\documentclass[fonts, doublespace]{icst}
\usepackage{moreverb}
\usepackage[breaklinks,colorlinks,bookmarksopen,bookmarksnumbered,linkcolor=ICSTblue,citecolor=ICSTblue,urlcolor=ICSTblue]{hyperref}
\usepackage{breakurl}
\usepackage{doi}
\usepackage{amssymb}
\usepackage{algorithm}
\usepackage{algpseudocode}
\usepackage{graphicx}
\usepackage{float}
\usepackage[justification=centering]{caption}

\newcommand\BibTeX{{\rmfamily B\kern-.05em \textsc{i\kern-.025em b}\kern-.08em
T\kern-.1667em\lower.7ex\hbox{E}\kern-.125emX}}

\def\volumeyear{2013}

\begin{document}

\runningheads{S. E. Mahmoodi, K. P. Subbalakshmi, et al.}{Sensing and Resource Allocation in a Cognitive Relay}

\title{Joint Spectrum Sensing and Resource Allocation for OFDM-based Transmission with a Cognitive Relay}

\author{S. Eman Mahmoodi\affil{1}\fnoteref{1}, K.P. Subbalakshmi\affil{1}, R. Chandramouli\affil{1}, Bahman Abolhassani\affil{2}}

\address{\affilnum{1}Department of Electrical and Computer Engineering, Stevens Institute of Technology, Hoboken, NJ, USA\\
\affilnum{2}Department of Electrical Engineering, Iran University of Science and Technology, Tehran, Iran}

\abstract{In this paper, we investigate the joint spectrum sensing and resource allocation problem to maximize throughput capacity of an OFDM-based cognitive radio link with a cognitive relay. By applying a cognitive relay that uses decode and forward (D\&F), we achieve more reliable communications, generating less interference (by needing less transmit power) and more diversity gain. In order to account for imperfections in spectrum sensing, the proposed schemes jointly modify energy detector thresholds and allocates transmit powers to all cognitive radio (CR) subcarriers, while simultaneously assigning subcarrier pairs for secondary users (SU) and the cognitive relay. This problem is cast as a constrained optimization problem with constraints on (1) interference introduced by the SU and the cognitive relay to the PUs; (2) miss-detection and false alarm probabilities and (3) subcarrier pairing for transmission on the SU transmitter and the cognitive relay and (4) minimum Quality of Service (QoS) for each CR subcarrier. We propose one optimal and two sub-optimal schemes all of which are compared to other schemes in the literature. Simulation results show that the proposed schemes achieve significantly higher throughput than other schemes in the literature for different relay situations.}


\fnotetext[1]{Corresponding author. \email{smahmood@stevens.edu}}

\maketitle

\section{Introduction}
Cognitive radio networks (CRNs) have been envisioned to provide efficient utilization of spectra for the secondary (unlicensed) users(SUs) without affecting the performance of the primary users (PUs) who are the primary licensees of the spectrum [1]. \emph{Orthogonal Frequency Division Multiplexing} (OFDM) is considered the most appropriate modulation scheme for secondary users [2], [3]. Power allocation for capacity maximization of an OFDM based CRN, while keeping the interference from the cognitive radio (CR) to the PUs below a given threshold was considered in [4]. In [5], the mutual interference power between PUs and the CR was calculated for power allocation, assuming OFDM for the PU transmission as well. Also in [6], a new opportunistic scheme for resource allocation was expressed based on interference cancellation. In our previous work [7], we derived the transmit power allocations without the assumptions of perfect knowledge of interference introduced by the PUs to the SUs. Also several prior works in this area, [4-7], assumed perfect channel sensing, which also is not guaranteed in practice. In [8], by adjusting the sensing threshold, joint sensing and resource allocation is performed. There is a trade-off between achieving maximum throughput of the CRN and guaranteeing the quality-of-service (QoS) of the PUs. By accessing more OFDM subchannels, the CRN obtains higher throughput and also introduces interferences caused by the miss-detection. So, joint resource allocation and sensing thresholds selection is needed to achieve optimum performance. However, [8] assumed that both PUs and the CR use OFDM and that the transmit power levels for spectrum sensing at each of the OFDM subchannels are known. In this paper, we allocate transmit power under more practical considerations, where the sensing mechanisms may lead to false alarms and miss probabilities. 

Decode and Forward (D\&F) relaying has been shown to achieve more reliable transmissions even at lower transmit power levels, thereby allowing for less interference to the PUs [9]. In this paper, we show that by applying D\&F relay, we can achieve higher precision in spectrum sensing, and we call such a relay \textit{cognitive relay}. Here, spectrum sensing is performed by the cognitive relay and the sensing thresholds are updated in the first time slot transmission [10]. A cognitive relay network under fading channels has been proposed [11], where better spectrum opportunity utilization has been achieved via higher diversity gain (and therefore higher system complexity) although under the assumptions of perfect spectrum sensing results. A two pronged approach, where imperfect soft sensing and a sensing update, has been used to alleviate the problem of sensing errors in [12] and [13]. In [12], the SU adapts the transmit power based on the detection and false alarm probabilities. In our previous work [13], we maximized throughput of the CRN by considering constraints on interference and total power for multiuser CRN under imperfect sensing conditions characterized by false alarm probabilities. 

Management of CR subcarriers is another problem in the cognitive relay transmission. The cognitive relay can decode and forward data on the same CR subcarrier, used by the secondary node [14-16]. However, a more efficient solution is to reconsider throughput capacity maximization of the CRN by CR subcarrier allocation for the cognitive relay. This means that the cognitive relay may not apply the same CR subcarrier as the SU transmitter (TX) [17]. Thus, two OFDM subcarriers are paired in the first and second time slot during the transmission by the SU TX and the cognitive relay, respectively.

The goal of this paper is to maximize the throughput capacity of the CRN, by jointly allocating transmit power levels, sensing spectrum (updating the values of energy detector thresholds), and pairing CR subcarriers to the SU TX and cognitive relay. Constraints of this optimization problem are: keeping the total interference introduced by the SU and cognitive relay below a given threshold, keeping miss-detection and false alarm probabilities in each CR subcarrier below a specified threshold, considering subcarrier pairing for subcarrier pairs of the SU and cognitive relay, and providing a lower bound on transmit power levels for CR subcarriers. First, convexity of this optimization problem is verified. Then, using Lagrange formulation, constraints' duality [18], taking subgradient method [19] and applying a greedy method, this problem will be solved. We will then propose low complexity, sub-optimal schemes and evaluate them. 

The rest of this paper is organized as follows. Section 2 describes the system model. Section 3 formulates the optimization problem by interference modeling. Section 4 proposes the main scheme for joint power allocation, spectrum sensing, and subcarrier pairing. Section 5 introduces the suboptimal and other schemes comparable to the main proposed scheme. Section 6 discusses simulation results by comparing five schemes with each other. Finally, Section 7 concludes the paper and the expected future works are expressed.

\section{System Model}
In this paper, we consider a two time slot, D\&F relay CRN consisting of a SU transceiver, and a cognitive relay, both of which coexist with a number of PUs. In addition to relaying transmitted data streams, the cognitive relay also performs wideband spectrum sensing with $N$ energy detection thresholds for the $N$ OFDM subcarriers. Fig. 1 depicts the system model. In the first time slot, we observe that the SU TX sends data on subcarrier $i$, while the cognitive relay and SU receiver (RX) receives the data. Then, the cognitive relay, which employs D\&F, transmits decoded data in the second time slot on subcarrier $j$, while the SU RX receives data. Since data transmission of the SU TX and cognitive relay occurs in different time slots, there is no interference between them. The SU RX applies \textit{Maximum Ratio Combining} (MRC) to obtain data by exploiting spatial diversity. The cognitive relay must decide which subcarrier to use for data transmission based on its channel sensing results. So, the cognitive relay may not transmit data in the same subcarrier that the SU TX used. We assume the time slots are of the same duration and the coherence time of channels are much longer than the time between CSI updates, hence channel gains ($h_{m}^{(ab)} $in subcarrier $m$ between TX $a$ and RX $b$ in Fig. 1) do not change while data transmission period once it is measured. Also, we assume that the sensing periods are smaller than the switching time of the PUs' activities.

In this paper, we improve the reliability of spectrum sensing results along with power allocation and subcarrier pairing. To perform wideband spectrum sensing, energy detectors are used for each CR subcarrier. Energy detection thresholds ($\lambda_{1},\lambda_{2},\cdots ,\lambda_{N}$) are applied on each of the $N$ CR subcarriers for spectrum sensing [20]. Let the fading channel gain between the primary TX and cognitive relay in subcarrier $i$ be $h_{i}^{(pr)}$. As shown in [20] based on Neyman-Pearson criterion, the probabilities of false alarm and detection in subcarrier $i$ are respectively given by $p_{f} (\lambda_{i} )=Q(\frac{\lambda_{i} -M\sigma_{n}^{2} }{\sqrt{2M} \sigma_{n}^{2} } )$ and $p_{d} (\lambda_{i} )=Q(\frac{\lambda_{i} -M(\sigma_{n}^{2} +|h_{i}^{(pr)}|^2)}{\sqrt{2M\sigma_{n}^{2} (\sigma_{n}^{2} +2|h_{i}^{(pr)} |^{2} )} } )$, where $M$ denotes the number of previous received samples taken in the cognitive relay for spectrum sensing and $\sigma_{n}^{2} $ represents the variance of Additive White Gaussian Noise (AWGN) channel.

\begin{figure}
	\centering
		\includegraphics[width=0.5\textwidth]{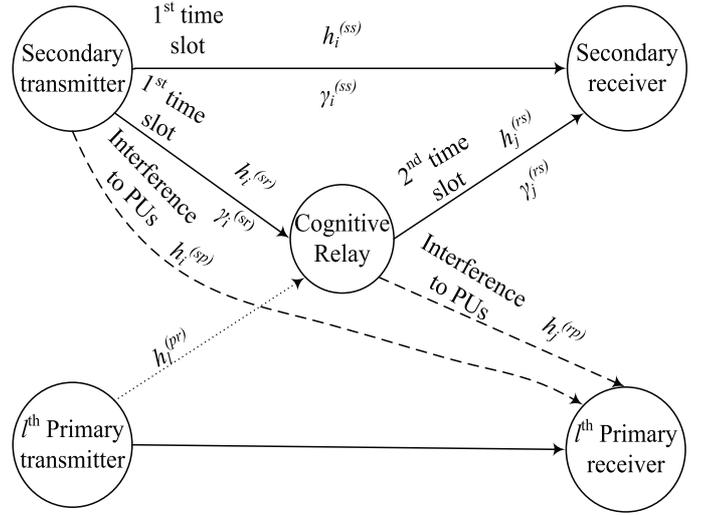}
	\caption{\footnotesize{System Model of the CRN in Two Time Slot Transmission.}}
\end{figure}
Let the ratio of channel gain to interference power introduced by $L$ OFDM subchannels of PUs in subcarrier $m$, from TX $a$, at RX $b$, be $\gamma_{m}^{(ab)}$. Then $\gamma_{i}^{(sr)} =\frac{|h_{i}^{(sr)} |^{2} }{\sigma_{s}^{2} +\sum_{l=1}^{L}J_{i}^{(pr),(l)}  } $, $\gamma_{i}^{(ss)} =\frac{|h_{i}^{(ss)} |^{2} }{\sigma_{s}^{2} +\sum_{l=1}^{L}J_{i}^{(ps),(l)}  }$, $\gamma_{j}^{(rs)} =\frac{|h_{j}^{(rs)} |^{2} }{\sigma_{s}^{2} +\sum_{l=1}^{L}J_{j}^{(ps),(l)}  } $, where we assume the same AWGN variance $\sigma_{s}^{2}$ for all three links and $L$ denotes the number of OFDM subchannels occupied by PUs. The interference power introduced by PU's subchannel $l$ to the CR subcarrier $i$ is given by $J_{i}^{(ps),(l)} =|h_{l}^{(ps)} |^{2} \int_{\Delta_{i,l} -\frac{\Delta f}{2} }^{\Delta_{i,l} +\frac{\Delta f}{2} }\Phi_{PER} (w)dw $, [21] where the expected periodogram is $\Phi_{PER} (w)=\frac{1} {2\pi K} \int_{-\pi }^{+\pi }\phi_{{PU} }^{(l)} (e^{jw} )(\frac{\sin ((w-\nu )k/2)}{\sin ((w-\nu )/2)} )^{2} d\nu$, $K$ is the Fast Fourier Transform (FFT) size of the periodogram, $w$ is the normalized frequency, $\Delta_{i,l}$ is the spectral distance between CR subcarrier $i$ and PU's subchannel $l$, and $\phi_{PU}^{(l)} $ is the power spectrum density (PSD) of PU's subchannel $l$. The achievable weighted rate with an ideal coding scheme, for subcarrier pair ${\rm SP}(i,j)$ can be expressed as $R_{i,j} =\frac{\Delta f\rho_{i} }{2} \log_{2} (1+\gamma_{i,j} P_{i,j} )$, where $\Delta f$ and $\rho_{i}$ are the frequency space of OFDM subcarriers and weight factor for subcarrier $i$, respectively [14]. The weight factors are taken into consideration to reflect distinct QoS requirements for each subcarrier in the CRN. Note that $\rho_{i} $ is dependent only on CR subcarrier $i$, which is used by the SU TX. $\gamma_{i,j}$ is the equivalent channel gain of ${\rm SP}(i,j)$ and $P_{i,j}$ is the sum of two powers: the transmit power of SU TX in subcarrier $i$ when the cognitive relay is transmitting in subcarrier $j$ ($P_{i,j}^{(1)} $) and the transmit power of the cognitive relay in subcarrier $j$ when the SU TX is transmitting in subcarrier $i$ ($P_{i,j}^{(2)} $), $P_{i,j} =P_{i,j}^{(1)} +P_{i,j}^{(2)}$.

The communications in our model can either follow the direct mode or relay mode depending on whether the channel gains to noise ratio ($\gamma_{i,j}$s) is favorable (relay) or not (direct). As in [22], the power levels and equivalent channel gain to noise plus interference ratio for ${\rm SP}(i,j)$ in the relay link (if $\gamma_{i}^{(sr)} \ge \gamma_{i}^{(ss)} {\rm \& }\; \gamma_{j}^{(rs)} \ge \gamma_{i}^{(ss)}$) are formulated by
\begin{eqnarray} 
\label{GrindEQ__3_} 
\left\{\begin{array}{l} {P_{i,j}^{(1)} =\frac{\gamma_{j}^{(rs)} }{\gamma_{i}^{(sr)} +\gamma_{j}^{(rs)} -\gamma_{i}^{(ss)} } P_{i,j} ,}  \\ P_{i,j}^{(2)} =\frac{\gamma_{i}^{(sr)} -\gamma_{i}^{(ss)} }{\gamma_{i}^{(sr)} +\gamma_{j}^{(rs)} -\gamma_{i}^{(ss)} } P_{i,j} ,\\ 
\gamma_{i,j} =\frac{\gamma_{i}^{(sr)} \gamma_{j}^{(rs)} }{\gamma_{i}^{(sr)} +\gamma_{j}^{(rs)} -\gamma_{i}^{(ss)} } . 
\end{array}\right.  
\end{eqnarray}

Otherwise, for the direct link transmission, we have
\begin{eqnarray} 
\label{GrindEQ__4_} 
\left\{\begin{array}{l} {P_{i,j}^{(1)} =P_{i,j} ,} \\ {P_{i,j}^{(2)} =0,} \\ {\gamma _{i,j} =\gamma _{i}^{(ss)} .} \end{array}\right. \;  
\end{eqnarray} 

\section{Interference Modeling and Problem Formulation}
We define an objective function for the transmission link on ${\rm SP}(i,j)$ as, $\mathcal{A}_{i,j} \times \mathcal{B}_{i,j} $, that gives us a sense of both the throughput ($\mathcal{A}_{i,j}$) and the capacity ($\mathcal{B}_{i,j}$), where $\mathcal{A}_{i,j} =(1-p_{f} (\lambda_{i} ))(1-p_{f} (\lambda_{j} ))$and $\mathcal{B}_{i,j} =\frac{\Delta f\rho _{i} }{2} \log_{2} (1+\gamma_{i,j} P_{i,j})$. $\mathcal{A}_{i,j} $ increases as the false alarm decreases, just as we would expect the throughput of the system to do. So, in a sense $\mathcal{A}_{i,j}$ is related to the throughput. $\mathcal{B}_{i,j} $ is a measure of the capacity of the system in transmission of ${\rm SP}(i,j)$. Hence the objective function for ${\rm SP}(i,j)$ is defined as:
\begin{equation} \label{GrindEQ__5_} 
C_{i,j} =q_{i,j} (1-p_{f} (\lambda _{i} ))(1-p_{f} (\lambda _{j} ))\frac{\Delta f\rho _{i} }{2} \log _{2} (1+\gamma _{i,j} P_{i,j} ), 
\end{equation} 
where, $q_{i,j}$ is an indicator function taking on a value 1 when the ${\rm SP}(i,j)$ is used and zero otherwise. Note that Eqn. \eqref{GrindEQ__5_} has three sets of unknown parameters: the equivalent transmit power levels ($P_{i,j}$), subcarrier pairing indicators ($q_{i,j}$) and energy detector thresholds ($\lambda _{i} $),$\forall i,j=1,2,...,N$. The objective function is discrete because the selection of subcarrier pairs is discrete. In order to deal with this, we apply continuous relaxation to this binary constraint ($q_{i,j}$) by writing the objective function for ${\rm SP}(i,j)$ as 
\begin{equation} \label{GrindEQ__6_} 
C_{i,j} =q_{i,j} (1-p_{f} (\lambda_{i} ))(1-p_{f} (\lambda _{j} ))\frac{\Delta f\rho _{i} }{2} \log _{2} (1+\gamma _{i,j} \frac{P_{i,j} }{q_{i,j} } ), 
\end{equation} 
where the binary constraint changes to: $q_{i,j}\ge 0, i,j= 0,1,2,\dots ,N$ [23]. 

As illustrated in Fig. 1, the interference is introduced to the PUs by each secondary node transmission (either the $j^{th}$ subcarrier of the SU TX or the $i^{th}$ subcarrier of the cognitive relay). These interference power levels can be calculated as $P_{I,j}^{(s)} =\sum _{l=1}^{L}\sum _{i=1}^{N}P_{i,j}^{(1)} \phi _{i,s}^{(l)} $ and $P_{I,i}^{(r)} =\sum _{l=1}^{L}\sum _{j=1}^{N}P_{i,j}^{(2)} \phi _{j,r}^{(l)}$, respectively. The interference power spectral densities introduced by CR subcarrier ${i}$ on the primary subchannel \textit{l} is given by $\phi _{i,s}^{(l)} =|h_{l}^{(sp)} |^{2} T_{s} \int _{\Delta _{i,l} -B_{l} /2}^{\Delta _{i,l} +B_{l} /2}(\frac{\sin (\pi fT_{s} )}{\pi fT_{s} } )^{2}  df,$ $\phi _{i,r}^{(l)} =|h_{l}^{(rp)} |^{2} T_{s} \int _{\Delta _{i,l} -B_{l} /2}^{\Delta _{i,l} +B_{l} /2}(\frac{\sin (\pi fT_{s} )}{\pi fT_{s} } )^{2}  df$ can be calculated assuming ideal Nyquist transmitted pulse [21] and OFDM symbol duration, $T_s$.

One of the main constraints in this optimization problem is to keep the interference power introduced by SU and the cognitive relay below the specified interference power threshold $(P_{I}^{(th)}).$ So, we consider $P_{I}^{(s)} \le P_{I}^{(th)} {\rm \; }{\rm and}\; P_{I}^{(r)} \le P_{I}^{(th)} ,$which can be rewritten as a function of $P_{i,j}$ by
\begin{equation} \label{GrindEQ__10_} 
\left\{\begin{array}{c} {\sum _{j=1}^{N}\sum _{l=1}^{L}\{ \sum _{i=1}^{N}P_{i,j} \Phi _{i,j}^{s,(l)} \}    \le P_{I}^{(th)} ,{\rm \; \; \; \; \; \; \; \; \; \; \; \; \; \; \; \; \; \; \; \; \; \; \;}} \\{\sum_{i=1}^{N}\sum _{l=1}^{L}\{ \sum _{j=1}^{N}P_{i,j} \Phi _{i,j}^{r,(l)} \}    \le P_{I}^{(th)} .{\rm \; \; \; \; \; \; \; \; \; \; \; \; \; \; \; \; \; \; \; \; \; \; \; }} \end{array}\right.  
\end{equation} 
where
\begin{equation} \label{GrindEQ__11_} 
\left\{\begin{array}{c} {\Phi _{i,j}^{s,(l)} =\frac{\gamma _{j}^{(rs)} }{\gamma _{i}^{(sr)} +\gamma _{j}^{(rs)} -\gamma _{i}^{(ss)} } \phi _{i,s}^{(l)} ,} \\ {\Phi _{i,j}^{r,(l)} =\frac{\gamma _{i}^{(sr)} -\gamma _{i}^{(ss)} }{\gamma _{i}^{(sr)} +\gamma _{j}^{(rs)} -\gamma _{i}^{(ss)} } \phi _{j,r}^{(l)} .} \end{array}\right.  
\end{equation} 
The objective function, which is called $throughput capacity$ in this paper, is maximized over \textbf{P, q}, $\boldsymbol{\lambda}$:

\begin{equation} \label{GrindEQ__12_}
\begin{array}{l} 
\mathop{\max }\limits_{\textbf{P, q}, \boldsymbol{\lambda}} \sum _{i=1}^{N}\sum _{j=1}^{N}q_{i,j} \frac{\Delta f\rho _{i} }{2} (1-p_{f} (\lambda _{i} ))(1-p_{f} (\lambda _{j} ))\\{\rm \; \; \; \; \; \; \; \; \; \; \; \; \; \; \; \; \; \; \; \; \; \; \; \; \; \; \; } \times\log _{2} \left(1+\gamma _{i,j} \frac{P_{i,j} }{q_{i,j} } \right)  , 
\end{array} 
\end{equation}  

subject to the connections in Eqn \eqref{GrindEQ__10_} and:
\begin{equation} \label{GrindEQ__13_} 
1-p_{d} (\lambda _{j} )\le \alpha _{j} \; \; \; \; \forall j=1,2,..,N, 
\end{equation} 
\begin{equation} \label{GrindEQ__14_} 
p_{f} (\lambda _{j} )\le \beta _{j} \; \; \; \; \; \; \; \; \forall j=1,2,..,N, 
\end{equation} 
\begin{equation} \label{GrindEQ__15_} 
\sum _{i=1}^{N}q_{i,j} =1 \; \; \; \; \; \; \; \; \; \; \forall j=1,2,..,N, 
\end{equation} 
\begin{equation} \label{GrindEQ__16_} 
\sum _{j=1}^{N}q_{i,j} =1 \; \; \; \; \; \; \; \; \; \; \; \forall i=1,2,..,N, 
\end{equation} 
\begin{equation} \label{GrindEQ__17_} 
P_{i,j} \ge \nu _{i,j} \; \; \; \; \; \; \; \; \forall i,j=1,2,..,N,\forall P_{i,j} \ne 0, 
\end{equation} 
\begin{equation} \label{GrindEQ__18_} 
q_{i,j} \ge 0\; \; \; \; \; \; \; \; \; \; \; \; \forall i,j=1,2,..,N, 
\end{equation} 
\begin{equation} \label{GrindEQ__19_} 
\lambda _{j} \ge 0\; \; \; \; \; \; \; \; \; \; \; \; \; \; \; \forall j=1,2,..,N. 
\end{equation} 
In Eqn \eqref{GrindEQ__12_}, \textbf{P} and \textbf{q} are ($N\times N$) matrices with entries ($P_{i,j}$,  $q_{i,j}$, $\forall i,j=1,2,...,N$) and  $\lambda =[\lambda_{1},\lambda_{2}, ...,\lambda_{N}]$. The constraints in Eqns 
\begin{figure}
	\centering
		\includegraphics[width=0.5\textwidth]{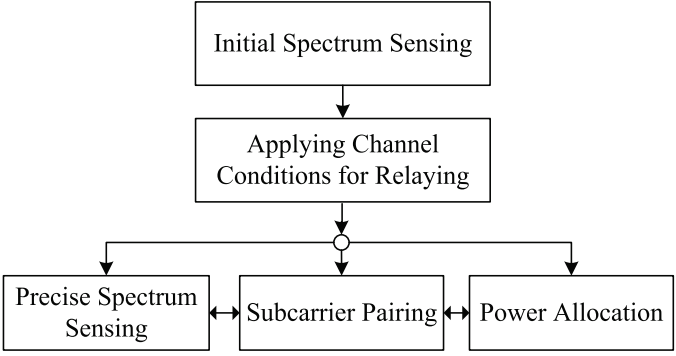}
		\caption{\footnotesize{Flowchart of the proposed scheme.}}
\end{figure}
\noindent \eqref{GrindEQ__13_} and \eqref{GrindEQ__14_} show that the miss-detection and false alarm probabilities for each CR subcarrier by the assigned energy detector thresholds should be kept below given thresholds to provide both efficient performance of the CRN and convexity of the problem. Also, Eqns \eqref{GrindEQ__15_} and \eqref{GrindEQ__16_} respectively denote that for each of the \textit{N} number of CR subcarriers allocated to the cognitive relay, only one CR subcarrier can be used for the SU TX and for each of CR subcarriers allocated to the SU TX, only one CR subcarrier can be used for the cognitive relay. We know that each of \textit{P${}_{i,j}$}s should be nonnegative, but the constraint in Eqn \eqref{GrindEQ__17_} is used to the convexity of the problem. The proof of the convexity of this optimization problem with all its constraints is given in the Appendix.

\section{ Joint Subcarrier Pairing, Power Allocation and Spectrum Sensing}
Our objective is to maximize throughput capacity (given by Eqn \eqref{GrindEQ__12_}) of the cognitive relay network by joint power allocation on the qualified subcarrier pairing and spectrum sensing. Fig. 2 illustrates a flowchart of the communication procedure from the view point of TX to provide an efficient transmission. First, the CR obtains an initial spectrum sensing output from the cognitive relay to know which OFDM subchannels are idle. Then, using the channel information and applying Eqns \eqref{GrindEQ__3_} and \eqref{GrindEQ__4_}, the decision on whether to take a direct link or a relay link is made. The optimization problem is then solved for the optimum transmit power levels, energy detector thresholds and subcarrier pairs. By applying Karush-Kahn-Tucker (KKT) conditions in the convex optimization, taking subgradient method in the dual problem and using an iterative algorithm, the optimization problem can be solved.

\subsection{ The Dual Problem }
We first obtain Lagrangian formulation of the objective function given by Eqn \eqref{GrindEQ__12_} and normalized to the bandwidth as
\begin{equation} \label{GrindEQ__20_} 
\begin{array}{l} 
C(\textbf{P, q},\boldsymbol{\lambda} ,\eta ,\kappa ,\boldsymbol{\tau ,\mu ,\delta} )=\frac{1}{2} \sum _{i=1}^{N}\sum _{j=1}^{N}\{ \rho _{i} q_{i,j} (1-p_{f} (\lambda _{j} ))\\{{\rm \; \; \; \; \; \; \; \; \; \; \; \; \; \; \; \; \; \; \; \; \; \; \; \; \; \; \; }\times(1-p_{f} (\lambda _{i} ))  } \log _{2} (1+\frac{\gamma _{i,j} }{q_{i,j} } P_{i,j} )\} \\ {{\rm \; \; \;  \; \; \; \; \; \; \; \; \; \; \; \; \; \; \; \; \; \; \; \; \; \; \; \; } +\eta (P_{I}^{(th)} -\sum _{l=1}^{L}\sum _{i=1}^{N}\sum _{j=1}^{N}P_{i,j} \Phi _{i,j}^{s,(l)}    )} \\ {\rm \; \; \; \; \; \; \; \; \; \; \; \; \; \; \; \; \; \; \; \; \; \; \; \; \; \; }+\kappa (P_{I}^{(th)} -\sum _{l=1}^{L}\sum _{i=1}^{N}\sum _{j=1}^{N}P_{i,j} \Phi _{i,j}^{r,(l)} \ )\\{{\rm \;  \; \; \; \; \; \; \; \; \; \; \; \; \; \; \; \; \; \; \; \; \; \; \; \; \; }-\sum _{j=1}^{N}\tau _{j} (1-\sum _{i=1}^{N}q_{i,j}  ) }+\sum _{i=1}^{N}\mu _{i} \\ {{\rm  \; \; \; \; \; \; \; \; \; \; \; \; \; \; \; \; \; \; \; \; \; \; \; \; \; \; }\times (\beta -p_{f} (\lambda _{i} )) +\sum _{i=1}^{N}\delta _{i} (p_{d} (\lambda _{i} )-1+\alpha ) .{\rm \; }} \end{array} 
\end{equation}
where $\eta , \kappa$ are Lagrange multipliers for the constraint in Eqn \eqref{GrindEQ__10_} and $\boldsymbol{\tau ,\mu ,\delta} $ are Lagrange vectors for the constraints in Eqns \eqref{GrindEQ__15_}, \eqref{GrindEQ__14_} and \eqref{GrindEQ__13_}, respectively. For convenience, we assume the same threshold for miss detection and fales alarm probabilities in each subcarrier ($\alpha_j=\alpha$ and $\beta_j=\beta$, $\forall j$). The constraint in Eqn \eqref{GrindEQ__16_} is studied in the proposed algorithm in Subsection 4.2, which simultaneously considers the satisfaction of this constraint with the other constraints. To maximize the throughput capacity for each ${\rm SP}(i,j)$ given by Eqn \eqref{GrindEQ__6_}, with respect to transmit power levels, we equate $\frac{\partial C_{i,j} }{\partial P_{i,j} } $=0:
\begin{equation} \label{GrindEQ__21_} 
\begin{array}{l}
\frac{\partial C_{i,j} }{\partial P_{i,j} } =\frac{\frac{\rho _{i} \gamma _{i,j} }{2} (1-p_{f} (\lambda _{j} ))(1-p_{f} (\lambda _{i} ))}{1+(\frac{\gamma _{i,j} }{q_{i,j} } )P_{i,j} } -\eta (\sum _{l=1}^{L}\Phi _{i,j}^{s,(l)}  )\\\\ {{\rm \; \; \; \; \; \; \; \; \; \; }-\kappa (\sum _{l=1}^{L}\Phi _{i,j}^{r,(l)}  )=0. 
}
\end{array}
\end{equation} 
We see that effect of the constraint in Eqn \eqref{GrindEQ__16_} is not considered in this formulation and the objective function is maximized by optimizing over two Lagrange variables. We solve the dual optimization problem with respect to Eqn \eqref{GrindEQ__17_}, and simplifying Eqn \eqref{GrindEQ__21_} to get
\begin{equation} \label{GrindEQ__22_} 
P_{i,j}^{*} =q_{i,j} [\nu _{i,j} ,\frac{\frac{\rho _{i} }{2} (1-p_{f} (\lambda _{i} ))(1-p_{f} (\lambda _{j} ))}{\eta (\sum _{l=1}^{L}\Phi _{i,j}^{s,(l)}  )+\kappa (\sum _{l=1}^{L}\Phi _{i,j}^{r,(l)}  )} -\frac{1}{\gamma _{i,j} } ]^{+} , 
\end{equation} 
where [${x,y}$]${}^{+}$=max(${x,y}$). Since the power is dependent on the energy detector thresholds and the subcarrier pair selection, we jointly optimize these parameters. Also, according to Eqn \eqref{GrindEQ__17_}, the power may be zero for some subcarrier pairs. From Eqn \eqref{GrindEQ__22_}, we see that this occurs when \textit{q${}_{i,j}$}=0.

For spectrum sensing, we achieve the values of energy detector thresholds over \textit{N} CR subcarriers, which are used either by SU TX or the cognitive relay in transmission. So, the objective function given by Eqn \eqref{GrindEQ__5_} is considered for one set of $N$ CR subcarriers (for $i=1,...,N$). Then, simplified formulation for Eqn \eqref{GrindEQ__20_} is written as
\begin{equation} \label{GrindEQ__23_} 
\begin{array}{l} 
\breve{C}=\frac{1}{2} \sum _{i=1}^{N}(1-p_{f} (\lambda _{i} ))\log _{2} (1+\gamma _{i,i} P_{i,i}^{*} ) +\eta (P_{I}^{(th)} \\ {{\rm \;  \;  \; \; } -\sum _{l=1}^{L}\sum _{i=1}^{N}\sum _{j=1}^{N}P_{i,j}^{*} \Phi _{i,j}^{s,(l)}    )+}\kappa (P_{I}^{(th)}\\ {{\rm \;  \;  \; \; }  -\sum _{l=1}^{L}\sum _{i=1}^{N}\sum _{j=1}^{N}P_{i,j}^{*} \Phi _{i,j}^{r,(l)}    ) +\sum _{j=1}^{N}\tau _{j} (1-\sum _{i=1}^{N}q_{i,j}  ) }\\ {{\rm \;  \;  \; \; }  +\sum _{i=1}^{N}\mu _{i} (\beta -p_{f} (\lambda _{i} ))+\sum _{i=1}^{N}\delta _{i} (p_{d} (\lambda _{i} )-1+\alpha ) .{\rm \; \; \; \; \; \; }} 
\end{array} 
\end{equation}
Then, to obtain $\lambda_{i}$, we have 
\begin{equation} \label{GrindEQ__24_} 
\frac{\partial \breve{C}_{i} }{\partial \lambda _{i} } =\frac{-(\mu _{i} +\tilde{R}_{i} )}{\sqrt{4\pi M} \sigma _{n}^{2} } e^{-\frac{x_{i}^{2} }{2} } +\frac{\delta _{i} }{\sqrt{4\pi M\sigma _{n}^{2} (\sigma _{n}^{2} +2E(|s_{i} h_{i}^{(pr)} |^{2} )} } e^{-\frac{y_{i}^{2} }{2} } , 
\end{equation} 
where $x_{i} =\frac{\lambda _{i} -M\sigma _{n}^{2} }{\sqrt{2M} \sigma _{n}^{2} } {\rm ,\; }y_{i} =\frac{\lambda _{i} -M(\sigma _{n}^{2} +E(|s_{i} h_{i}^{(pr)} |^{2} )}{\sqrt{2M\sigma _{n}^{2} (\sigma _{n}^{2} +2E(|s_{i} h_{i}^{(pr)} |^{2} )} } $ and$\tilde{R}_{i} $ is the normalized rate of CR subcarrier \textit{i};
\begin{equation} \label{GrindEQ__25_} 
\begin{array}{l}
\tilde{R}_{i} \mathop{=}\limits^{\Delta } \frac{1}{2} \log _{2} (1\\ {{\rm \;  \;  \; \;\;  \;  \; \;\;  \;  \; \;\;  \;  \; \; }+\gamma _{i,i} [\frac{e-1}{\gamma _{i,i}^{} } ,\frac{1}{2\eta \sum _{l=1}^{L}\Phi _{i,i}^{s,(l)} + 2\kappa \sum _{l=1}^{L}\Phi _{i,i}^{r,(l)}  } -\frac{1}{\gamma _{i,i}^{} } ]^{+} ). 
}
\end{array}
\end{equation}
 By setting Eqn \eqref{GrindEQ__24_} to zero, we have
\begin{equation} \label{GrindEQ__26_} 
\frac{y_{i}^{2} -x_{i}^{2} }{2} =\ln (\frac{\delta _{i} }{\sqrt{4\pi M\sigma _{n}^{2} (\sigma _{n}^{2} +2E(|s_{i} h_{i}^{(pr)} |^{2} )} } )-\ln (\frac{\mu _{i} +\tilde{R}_{i} }{\sqrt{4\pi M} \sigma _{n}^{2} } ), 
\end{equation} 
and Eqn \eqref{GrindEQ__26_} simplifies to
\begin{equation} \label{GrindEQ__27_} 
\begin{array}{l}
\frac{\lambda _{i} -M(\sigma _{n}^{2} +E(|s_{i} h_{i}^{(pr)} |^{2} )}{\sqrt{2M\sigma _{n}^{2} (\sigma _{n}^{2} +2E(|s_{i} h_{i}^{(pr)} |^{2} )} } )^{2} -(\frac{\lambda _{i} -M\sigma _{n}^{2} }{\sqrt{2M} \sigma _{n}^{2} } )^{2} \;  \;  \; \;=\\ {{\rm \;  \;  \; \; } 2\ln \frac{\sigma _{n} \xi _{i} }{\sqrt{\sigma _{n}^{2} +2E(|s_{i} h_{i}^{(pr)} |^{2} } } ), }
\end{array}
\end{equation}
where $\xi _{i} =\frac{\delta _{i} }{\mu _{i} +\tilde{R}_{i} } $. By solving Eqn \eqref{GrindEQ__19_}, two values for $\lambda_{i}$ are obtained and only one of them is nonnegative (Constraint (19)) and acceptable, which is given by
\begin{equation} \label{GrindEQ__28_} \begin{array}{l}
\lambda _{i}^{\circ } =\frac{M\sigma _{n}^{2} }{2|s_{i} h_{i}^{(pr)} |} \{ (\sigma _{n}^{4} +2\sigma _{n}^{2} |s_{i} h_{i}^{(pr)} |^{2} )\\ {{\rm \;  \;  \; \; } \times [\frac{M|s_{i} h_{i}^{(pr)} |^{2} -8\sigma _{n}^{2} \ln (\frac{\sigma _{n} \xi _{i} }{\sqrt{\sigma _{n}^{2} +2|s_{i} h_{i}^{(pr)} |^{2} } } )}{M\sigma _{n}^{6} (\sigma _{n}^{2} +2|s_{i} h_{i}^{(pr)} |^{2} )} ]^{1/2} +|s_{i} h_{i}^{(pr)} |\} , }\end{array}
\end{equation} 

\noindent where the two Lagrange multipliers $\delta_{i}$ and $\mu_{i}$ are combined and given by $\xi _{i} =\frac{\delta _{i} }{\mu _{i} +\tilde{R}_{i} } .$We set $\mu_{i}$=0 in Eqn \eqref{GrindEQ__28_}, and by setting a lower bound to satisfy the false alarm constraint given by Eqn \eqref{GrindEQ__14_}, we obtain   
\begin{equation} \label{GrindEQ__29_} 
\lambda _{i}^{*} =\max ((Q^{-1} (\beta _{i} )\sqrt{2M} +M)\sigma _{n}^{2} ,\lambda _{i}^{\circ } ). 
\end{equation} 

We now obtain the optimal subcarrier pair indicators ($q_{i,j}^{*} $s). By substituting the allocated transmit power levels given by Eqn \eqref{GrindEQ__22_} into Eqn \eqref{GrindEQ__20_}, we have
\begin{equation} \label{GrindEQ__30_} 
\begin{array}{l} 
C=\sum _{i=1}^{N}\sum _{j=1}^{N}q_{i,j} \Omega _{i,j}   +((\eta +\kappa )P_{I}^{(th)} {\rm +}\sum _{j=1}^{N}\tau _{j} \\ {\rm \;  \;  \; \; }  +\sum _{i=1}^{N}\mu _{i} (\beta -p_{f} (\lambda _{i} )) +\sum _{i=1}^{N}\delta _{i} (p_{d} (\lambda _{i} )-1+\alpha ), 
\end{array}
\end{equation} 
where
\begin{equation} \label{GrindEQ__31_} 
\begin{array}{l} {\Omega _{i,j} =\frac{\rho _{i} }{2} \log _{2} (1+\gamma _{i,j} [\nu _{i,j} ,\frac{\frac{\rho _{i} }{2} (1-p_{f} (\lambda _{i} ))(1-p_{f} (\lambda _{j} ))}{\eta \sum _{l=1}^{L}\Phi _{i,j}^{s,(l)}  +\kappa \sum _{l=1}^{L}\Phi _{i,j}^{r,(l)}  } -\frac{1}{\gamma _{i,j} } )]^{+} )} \\ {{\rm \; \; \; \; \; \; \; }-\tau _{j} -\eta \sum _{l=1}^{L}\Phi _{i,j}^{s,(l)} [\nu _{i,j} ,\frac{\frac{\rho _{i} }{2} (1-p_{f} (\lambda _{i} ))(1-p_{f} (\lambda _{j} ))}{\eta \sum _{l=1}^{L}\Phi _{i,j}^{s,(l)}  +\kappa \sum _{l=1}^{L}\Phi _{i,j}^{r,(l)}  } -\frac{1}{\gamma _{i,j} } ]^{+}  } \\ {{\rm \; \; \; \; \; \; \; }-\kappa \sum _{l=1}^{L}\Phi _{i,j}^{r,(l)} [\nu _{i,j} ,\frac{\frac{\rho _{i} }{2} (1-p_{f} (\lambda _{i} ))(1-p_{f} (\lambda _{j} ))}{\eta \sum _{l=1}^{L}\Phi _{i,j}^{s,(l)}  +\kappa \sum _{l=1}^{L}\Phi _{i,j}^{r,(l)}  } -\frac{1}{\gamma _{i,j} } ]^{+}  .} \end{array} 
\end{equation} 
As can be seen from Eqn \eqref{GrindEQ__30_}, to maximize throughput capacity of the CRN over subcarrier pairs, $\Omega_{i,j}$ (for ${i}=1,...,{N, j=1,...,N})$ should be maximized. So, \textbf{\textit{q}} is assigned as 
\begin{equation} \label{GrindEQ__32_} 
q_{i,j}^{*} =\left\{\begin{array}{cc} {1} & {i=\arg \; \max \Omega _{i,j} \; \; \; \; \forall i=1,2,...,N,} \\ {0} & {{\rm otherwise}\; \; \; \; \; \; \; \; \; \; \; \; \forall j=1,2,...,N.} \end{array}\right.  
\end{equation} 
However, the subcarrier pairs obtained by Eqn \eqref{GrindEQ__32_} do not necessarily satisfy the constraint in Eqn \eqref{GrindEQ__16_}, in which only one CR subcarrier could be allocated for the cognitive relay in each pair. This problem will be solved in the following section by applying a greedy method in the proposed algorithm.

\subsection{ Algorithm Design}
In the previous subsection, all optimal transmit power levels, and subcarrier pairs were allocated and energy detector thresholds were obtained, by maximizing the throughput capacity given by Eqn \eqref{GrindEQ__12_} subject to constraints expressed in Eqns \eqref{GrindEQ__11_} and \eqref{GrindEQ__13_} to \eqref{GrindEQ__19_}, except \eqref{GrindEQ__16_}. The constraint in Eqn \eqref{GrindEQ__16_} specifies that each subcarrier should be allocated to one subcarrier pair for the cognitive relay transmission. In order to obtain optimal values for transmit power levels, and energy detector thresholds, as well as optimally pair subcarriers, and also satisfy Constraint \eqref{GrindEQ__16_} and exploit appropriate Lagrange multipliers, we propose an iterative algorithm, which is shown in by Algorithm 1.   

\begin{algorithm}
\caption{Proposed joint sensing, subcarrier pairing and power allocation.}
\begin{algorithmic} [1]
\State initialize Lagrange multipliers (for iteration: $k$=1)
\While {all Lagrange multipliers have higher errors than $\epsilon$} 
\State compute $\lambda_i$s by Eqn. \eqref{GrindEQ__29_},
\State compute $\Omega_{i,j}$s by Eqn. \eqref{GrindEQ__31_},
\State compute $q_{i,j}$s by Eqn. \eqref{GrindEQ__32_},
\State codify $q_{i,j}$s by Greedy method,
\State compute $P_{i,j}$s by Eqn. \eqref{GrindEQ__22_},
\State modify the values of Lagrange multipliers by
\State $\eta^{(k+1)}=\eta^{(k)}-\varepsilon_{1}^{(k)}(P_{I}^{(th)}-\sum_{j=1}^{N}\sum_{l=1}^{L}\{\sum_{i=1}^{N}P_{i,j}^{*(k)}\Phi_{i,j}^{s,(l)}\})$
\State $\kappa^{(k+1)}=\kappa^{(k)}-\varepsilon_{1}^{(k)}(P_{I}^{(th)}-\sum_{j=1}^{N}\sum_{l=1}^{L}\{\sum_{i=1}^{N}P_{i,j}^{*(k)}\Phi_{i,j}^{r,(l)}\})$
\State $\tau_{j}^{(k+1)}=\tau_{j}^{(k)}-\varepsilon_{2}^{(k)} (1-\sum_{i=1}^{N}q_{i,j}^{(k)}) \forall j=1,2,...,N,$
\State $\mu_{i}^{(k+1)}=\mu^{(k)}-\varepsilon_{3}^{(k)}[\beta - p_{f} (\lambda_{i}^{(k)})],$
\State $\delta_{i}^{(k+1)}=\delta^{(k)}-\varepsilon_{4}^{(k)} [p_{d} (\lambda_{i}^{(k)})-1+\alpha],$
\State $k=k+1$,
\EndWhile
\end{algorithmic}
\end{algorithm}
\begin{algorithm}
\caption{Greedy method to achieve \textbf{q} by satisfying constraint \eqref{GrindEQ__16_}.}
\begin{algorithmic} [1]
\State $t(j)=\sum _{i=1}^{N}q_{i,j}, \forall j=1,2,...,N$,

\For {$u=1:N$}
\If{$t(u)>1$} \State $b=arg\;\mathop{{\rm max}}\limits_{b|q_{b,u} =1}\Omega _{b,u}$ \EndIf
\While {$t(u)>1$}
\State $j{\rm =arg\; }\mathop{{\rm min}}\limits_{j|t(j)=0} {\rm \; }|\tau _{u} -\tau _{j}| { c{\rm =arg\; }\mathop{{\rm max}}\limits_{b|q_{c,u} =1,{\rm \; }c\ne b} {\rm \; }\Omega _{c,j}}$,
\State $t(u)=t(u)-1, t(j)=t(j)+1$,
\State $q(c,u)=0,q(c,j)=1$,
\EndWhile
\EndFor
\end{algorithmic}
\end{algorithm} 
In the proposed algorithm, we apply a greedy method similar to [17, 24] for satisfying Eqn \eqref{GrindEQ__16_}, which is shown by Algorithm 2. In summary, to maximize Eqn \eqref{GrindEQ__12_}, the greedy method finds the columns of \textbf{\textit{q }}with total value of more than one. Then, it places ones to the columns with sum value of zero, while the minimum deviation to Lagrange multiplier $\tau_j$ (for $j=1,2..,N$) is generated. Through this process, we achieve the minimum decrease in $\Omega_{i,j}$ (for ${i,j}=1,2..,N$) in the throughput capacity (Eqn \eqref{GrindEQ__12_}) of the CRN.
\subsection{ Complexity Analysis}
By applying the above algorithm, we jointly optimize the spectrum sensing, power allocation and subcarrier pairing in the cognitive relay network. We now compare the complexity of this algorithm with that of exhaustive search. Exhaustive search for subcarrier pairing has a complexity O(\textit{N}!), and for each fix subcarrier pair, the complexity of power allocations according to Eqn \eqref{GrindEQ__22_} and energy detector thresholds is O(\textit{NlogN}). Hence the complexity for exhaustive search is O(\textit{N.logN.N}!), which could be simplified to O(\textit{N.N}!). However, by applying the proposed algorithm, the complexity is O(\textit{N.logN}) per iteration. The number of iterations on the algorithm will depend on the value of \textit{$\varepsilon $}. Total complexity of this algorithm will be O(\textit{N}.log\textit{N})$\times$(number of iterations). Clearly this is much smaller than the exhaustive search method.

\section{ Suboptimal and Classical Schemes}
In this section, we propose a suboptimal scheme (in subsection 5.1) and introduce three classical schemes (in subsections 5.2, 5.3 and 5.4) for the throughput capacity maximization of the CRN, which is given by Eqn \eqref{GrindEQ__12_}, with limited interference threshold on PUs. All these schemes are compared using simulations.

\subsection{ Alternate Suboptimal Scheme}
In this scheme, we propose a suboptimal solution by simplifying the method of finding the Lagrange multipliers \textit{$\eta $ }and $\kappa $ to calculate energy detector thresholds. To do this, we consider a fixed Lagrange multiplier denoted by\textit{ $\eta $}\textbf{\textit{${}^{'}$}} and $\kappa '$ and then compute the energy detector thresholds for spectrum sensing. Using these thresholds, transmit power levels and subcarrier pairs are calculated. Finally, energy detector thresholds are modified with corresponding updated Lagrange multiplier for power allocation ($\eta$ and $\kappa$). Although this could cause a loss in the throughput capacity, the number of simultaneous variables to be calculated decreases and therefore the system complexity reduces. For exploiting the normalized rate, we first consider a uniform power distribution and according to Eqn \eqref{GrindEQ__10_}, we find $\eta'$ and $\kappa'$ for subcarrier pairs with nonzero powers. Then, by considering the boundary conditions in Eqn \eqref{GrindEQ__10_}, and replacing \textit{P${}_{i,j}$} with RHS of Eqn \eqref{GrindEQ__22_}, and assuming $p_{f}(\lambda_{i})= p_{f}(\lambda_{j})=0$, we write 
\begin{equation} \label{GrindEQ__33_}
\begin{array}{l}
\left\{\begin{array}{l} 
\sum _{j=1}^{N}\sum _{l=1}^{L}\{ \sum _{i=1}^{N}(\frac{\frac{\rho _{i} }{2} }{\eta '(\sum _{l=1}^{L}\Phi _{i,j}^{s,(l)}  )+\kappa '(\sum _{l=1}^{L}\Phi _{i,j}^{r,(l)}  )} -\frac{1}{\gamma _{i,j} } )\\ {{\rm \;  \;  \; \;\;  \;  \; \;\;  \;  \; \;  \;  \; \;\;  \;  \; \;\;  \;  \; \;\;\;  \;  \; \; } \times \Phi _{i,j}^{s,(l)} \}    =P_{I}^{(th)} ,} \\ \sum _{i=1}^{N}\sum _{l=1}^{L}\{ \sum _{j=1}^{N}(\frac{\frac{\rho _{i} }{2} }{\eta '(\sum _{l=1}^{L}\Phi _{i,j}^{s,(l)}  )+\kappa '(\sum _{l=1}^{L}\Phi _{i,j}^{r,(l)}  )} -\frac{1}{\gamma _{i,j} } )\\ {{\rm \;  \;  \; \;\;  \;  \; \;\;  \;  \; \;  \;  \; \;\;  \;  \; \;\;  \;  \; \;\;\;  \;  \; \; } \times\Phi _{i,j}^{r,(l)} \}    =P_{I}^{(th)} ,}
\end{array}\right. 
\end{array}
\end{equation} 
and from Eqn \eqref{GrindEQ__33_}, $\eta'$ and$\kappa'$ are obtained. As well, initial energy detector thresholds are calculated by this \textit{$\eta $}\textbf{\textit{${}^{' }$}} and $\kappa '$ based on Eqn \eqref{GrindEQ__29_}. Then, resource allocation is performed by one iteration of step 2 up to step 7, shown in Algorithm 1. Finally, modified energy detector thresholds are calculated by new updated \textit{$\eta $} and $\kappa $ obtained by one iteration of the algorithm.

\subsection{ Fixed Subcarrier Pairing (SCP) based Scheme }
Like [14], we consider a fixed assignment for subcarrier pairing in this scheme. In this scheme, the assigned subcarrier for the first time slot will be also used in the second time slot. So, we allocate power levels as
\begin{equation} \label{GrindEQ__34_} 
P_{i,i}^{*} =P_{i} =[\nu _{i,i} ,X_{i} -\frac{1}{\gamma _{i,i} } ]^{+} , 
\end{equation} 
where $X_i$ is the water level and it's based on the Lagrange multipliers $\eta$ and $\gamma$ and the false alarm probability over CR subcarrier $i$. So, it's given by
\begin{equation} \label{GrindEQ__35_} 
X_{i} =\frac{\frac{\rho _{i} }{2} (1-p_{f} (\lambda _{i} ))^{2} }{\eta (\sum _{l=1}^{L}\Phi _{i,j}^{s,(l)}  )+\kappa (\sum _{l=1}^{L}\Phi _{i,j}^{r,(l)}  )} . 
\end{equation}

\subsection{ Resource Allocation Based on Initial Spectrum Sensing (ISS)}
In this suboptimal scheme, we consider that spectrum sensing is performed by initial sensing results without updating the energy detector thresholds to obtain false alarm probabilities \textit{p${}_{f}$${}_{ }$}(\textit{$\lambda $${}_{i}$}), \textit{i}=1,2,...,\textit{N}, for each CR subcarrier based on [20]. Then, we jointly allocate transmit power levels, using Eqn \eqref{GrindEQ__22_}, and manage subcarrier pairs, using Eqn \eqref{GrindEQ__32_}.

\subsection{Classical Scheme Without use of the Cognitive Relay (WCR)}
Eventually, we compare the proposed schemes with a classical scheme, without use of the cognitive relay. In this scheme, transmit power is distributed according to the waterfilling scheme [25]. Based on Eqn \eqref{GrindEQ__10_}, and with regards to the given interference power threshold, the water levels are obtained. The power allocated to CR subcarrier \textit{i} is given by
\begin{equation} \label{GrindEQ__36_} 
P_{i}^{*} =[0,\frac{1}{\varpi } -\frac{1}{\gamma _{i}^{(ss)} } ]^{+} , 
\end{equation} 
and $\varpi $ is found by solving
\begin{equation} \label{GrindEQ__37_} 
\sum _{l=1}^{L}\sum _{i=1}^{N}[0,\frac{1}{\varpi } -\frac{1}{\gamma _{i}^{(ss)} } ]^{+}  \phi _{i,s}^{(l)}  =P_{I}^{(th)} . 
\end{equation} 
So, the interference constraint for the CRN is satisfied. By considering this constraint, the solution, which has been obtained in [4] is achieved. Note that spectrum sensing is performed like the previous scheme and false alarm and miss detection probabilities are obtained using [20].

\section{Simulation Results}
In this section, simulation results are presented and discussed. The performance of the main proposed scheme, introduced in Section 4, is compared with those of the other four schemes described in Section 5. We assume that the number of CR subcarriers (${N}$), PU occupied channels(${L}$), and previous received samples taken in the cognitive relay for spectrum sensing (${M}$) are 16, 48 and 32, respectively. In this system model, we consider an OFDM-based CR with cognitive relaying which coexists with three PUs. Also, the CRN uses the initial soft sensing results as the three PUs occupy 20, 12 and 16 subchannels. Note that in each run of the simulation, these three frequency bands of PUs are randomly distributed through OFDM subchannels. For the OFDM based CRN, we assume the OFDM frequency spacing ($\Delta{f}$) and symbol duration ($T_s$) to be 0.15625 MHz and 7 $\mu$S, respectively. AWGN variance in all the channels is assumed to be 10${}^{-5}$ Watts. The PU's transmit
\begin{figure}
	\centering
		\includegraphics[width=0.5\textwidth]{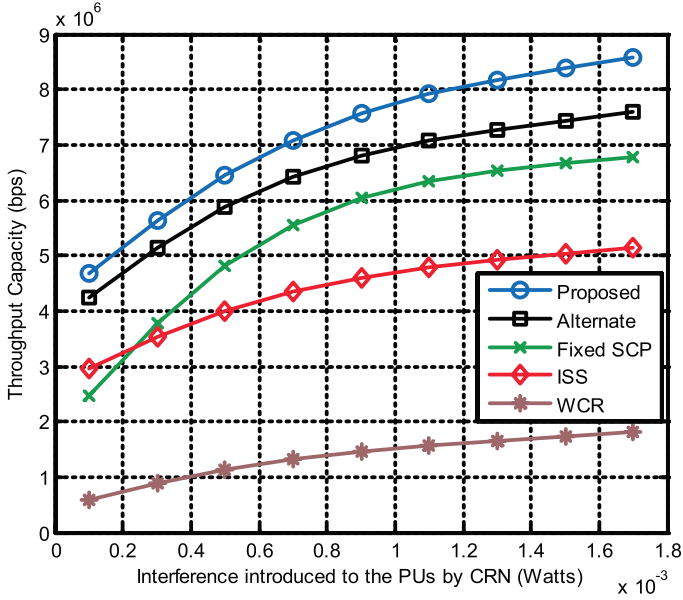}
\caption{\footnotesize{Throughput capacity of the CRN versus interference introduced to the PUs by several schemes with $E[|\gamma^{(sr)} |^2]=8$, $E[|\gamma^{(ss)} |^2]=3$ and $E[|\gamma^{(rs)} |^2]=8$.}}
\end{figure}
\noindent power ($P_{PU}$) is considered to be 5$\times$10${}^{-3}$ Watts. In the proposed algorithm, initial values of $\delta $ and \textit{$\tau $} for each CR subcarrier are randomly distributed between 0.01 and 2 with step size of 0.05/$\sqrt{k} $, where $k$ is the iteration index. The initial value for \textit{$\eta $} is randomly distributed between 100 and 200 and \textit{$\varepsilon $} is assumed to be 10${}^{-5}$.   

All the communication links are independent of each other in two time slots and their channel gains have Rayleigh distributions with given average channel gains. Except for three channels of $h^{(ss)}$, $h^{(rs)}$ and $h^{(sr)}$, which have different average channel gains through simulations, the ratio of average channel gains to noise plus interference in other links are set to 3. At first, we compare five schemes in a geographical situation where the cognitive relay is located between the SU TX and RX.
\begin{figure}
	\centering
		\includegraphics[width=0.5\textwidth]{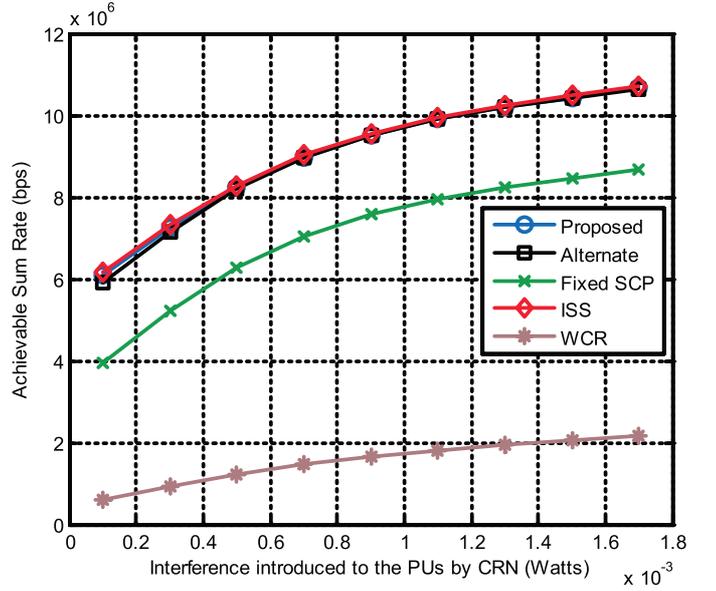}
\caption{\footnotesize{Maximum achievable sum rate of the CRN versus interference introduced to the PUs by several.}}
\end{figure}
\noindent So, we consider E[\textbar \textit{$\gamma $${}^{(sr)}$}\textbar ${}^{2}$]=8, E[\textbar \textit{$\gamma $${}^{(ss}$}\textbar ${}^{2}$]=3 and E[\textbar \textit{$\gamma $${}^{(rs)}$}\textbar ${}^{2}$]=8 in Figs. 3 to 7.

Fig. 3 shows the throughput capacity in bits/slot, which is defined in Eqn \eqref{GrindEQ__12_} in terms of interference power introduced to the PUs by the SU TX and the cognitive relay in Watts. In this figure, \textit{$\alpha $} and \textit{$\beta $} are assumed to be 0.2 and 0.3061, respectively. As we can see, the proposed scheme achieves the highest throughput for a given interference threshold. We also observe that Alternate scheme, where the sensing and the joint power allocation and subcarrier pairing are done in two excessive phases, performs close to the main proposed scheme. 

This scheme is able to achieve throughout capacities very close to the main proposed scheme, but with more iterations. However, decrease of the system complexity in this scheme is an advantage, but should be considered because it makes the precision of the approach lower. Note that by increase of iterations, Alternate scheme obtains the throughput close to the main proposed scheme but the system complexity increases. The fixed subcarrier pairing (SCP) based scheme and the scheme based on initial spectrum sensing (ISS) achieve less throughput than the proposed scheme and the proposed alternate scheme, but achieve more throughput than the scheme without use of cognitive relay (WCR). Note that false alarm probability in ISS and WCR based schemes are fixed and assumed to be 0.2. We observe that for the interference equal to 10${}^{-3}$ Watts, the main proposed, alternate, fixed SCP and ISS schemes have respectively 441\% , 363\%, 313\% and 213\% higher throughput capacity in comparison to WCR.

In Fig. 4, we plot the achievable total rate for the CRN versus interference power introduced to the PUs by the SU TX and the cognitive relay. As expected, since quality of spectrum sensing (false alarm probability) is not considered for the achievable total rate, performance of the main proposed scheme, alternate scheme, and ISS scheme are very close to each other. However, as can be understood from Fig. 3, spectrum sensing results are different in these schemes. We also observe that the fixed SCP and WCR achieve smaller rates compared to those of the other schemes.
\begin{figure}
	\centering
		\includegraphics[width=0.5\textwidth]{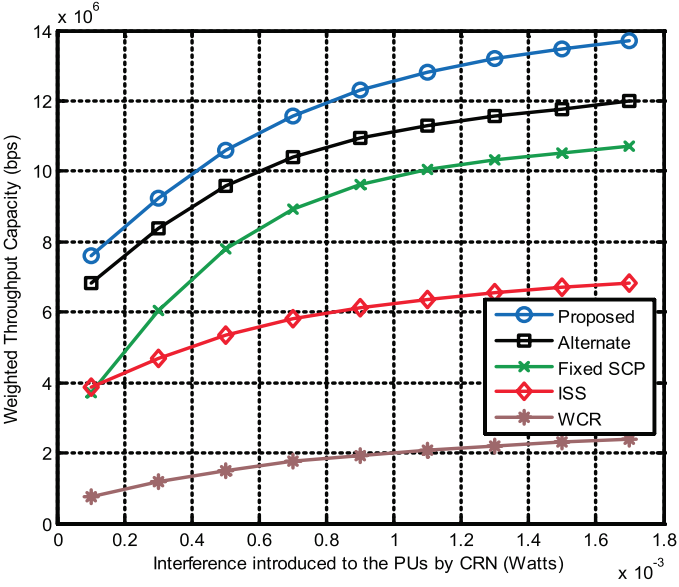}
	\caption{\footnotesize{Weighted throughput capacity of the CRN versus interference introduced to the PUs by several schemes.}}
\end{figure}

\begin{figure}
	\centering
		\includegraphics[width=0.5\textwidth]{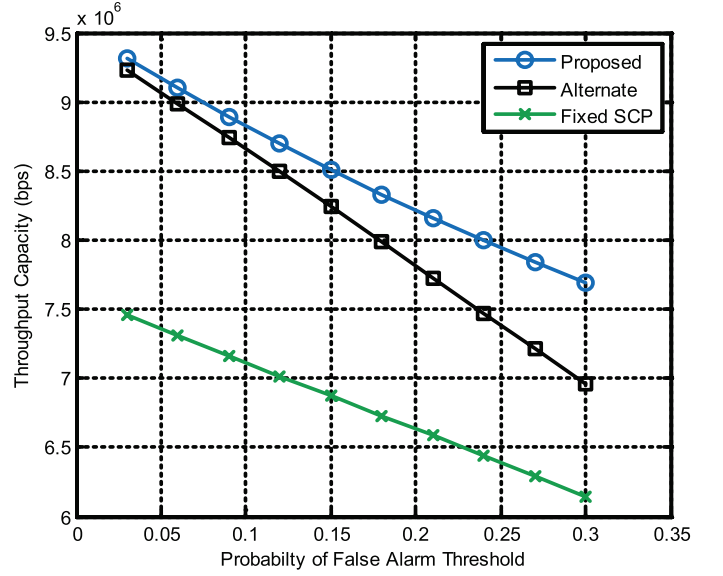}
	\caption{\footnotesize{Throughput capacity of the CRN versus probability of false alarm threshold by several schemes.}}
\end{figure}

In Fig. 5, we present the weighted throughput capacity given by Eqn \eqref{GrindEQ__12_} versus interference power introduced to the PUs by the SU TX and the cognitive relay. So far, we considered \textit{$\rho $${}_{i}$} (for all $i=1,2,\dots ,N$) as unity. However in this figure, we consider weighted rates such that \textit{$\rho $${}_{i}$}=1+(${i}$-1)/(${N}$-1),$\forall {i}$. This distribution is only an example for the weighted subcarriers in order to satisfy the QoS requirements in each one. In this condition, we observe that our main proposed scheme achieves higher throughput capacity for weighted CR subcarriers compared with those of the other schemes. Moreover, we observe that all of the schemes have simulation results similar to the simulation results shown in Fig. 3. 

In Fig. 6, we plot the throughput capacity in terms of probability of false alarm threshold. Here, \textit{I${}_{th}$} and \textit{$\alpha $} are assumed to be 10${}^{-3}$ Watts and 0.2, respectively. We see that the throughput of the CRN decreases by increasing the false alarm probability. 

Fig. 7 presents the total transmit power of the cognitive relay for all CR subcarriers versus the interference introduced to PUs by the SU TX and cognitive relay. We see that by subcarrier pairing, the main proposed, alternate and ISS schemes, allow for higher transmit power than that of the fixed SCP scheme for a given interference threshold. The fixed SCP does not take dynamic subcarrier pairing into account for the system. Therefore, this scheme allocates lower transmit power to the cognitive relay and achieves less total rate for the CRN.
\begin{figure}
	\centering
		\includegraphics[width=0.5\textwidth]{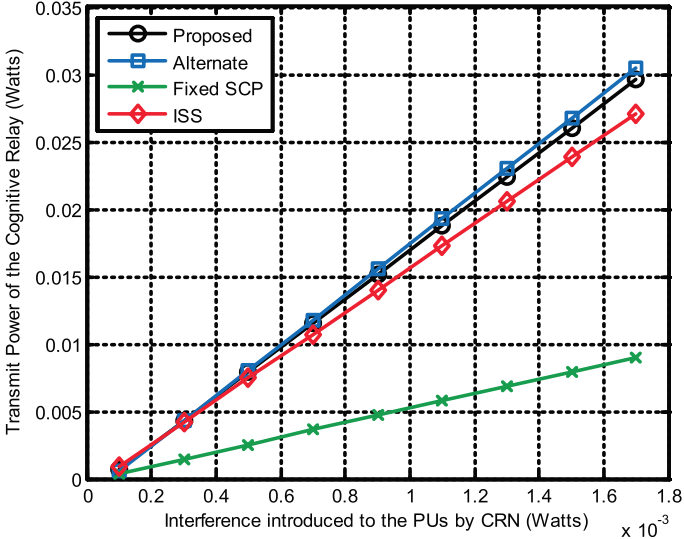}
	\caption{\footnotesize{Average transmit power of the cognitive relay versus interference introduced to the PUs by CRN by several schemes.}}
\end{figure}

\begin{figure}
	\centering
		\includegraphics[width=0.5\textwidth]{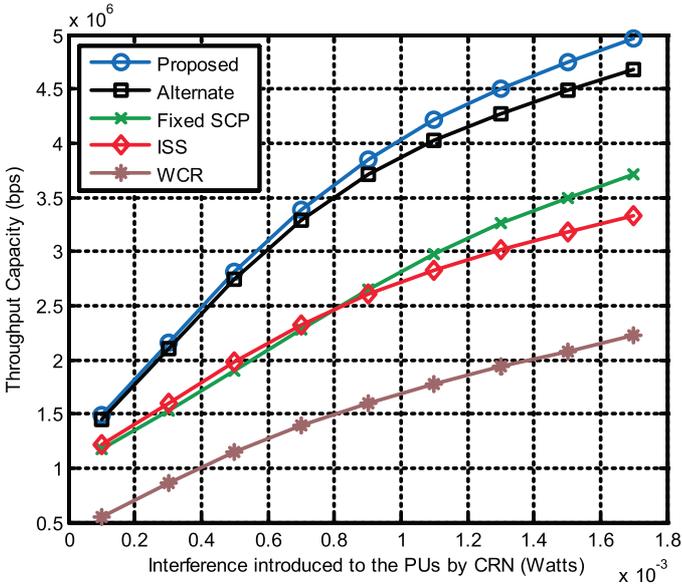}
\caption{\footnotesize{Throughput capacity of the CRN versus interference introduced to the PUs by several schemes with $E[|\gamma^{(sr)} |^2]=8$, $E[|\gamma^{(ss)} |^2]=3$ and $E[|\gamma^{(rs)} |^2]=8$ }}
\end{figure}
In Fig. 8, we consider the case where the cognitive relay is located close to the SU TX. So, we consider E[\textbar \textit{$\gamma $${}^{(sr)}$}\textbar ${}^{2}$]=8, E[\textbar \textit{$\gamma $${}^{(ss)}$}\textbar ${}^{2}$]=3 and E[\textbar \textit{$\gamma $${}^{(rs)}$}\textbar ${}^{2}$]=3. However, in Fig. 9,  we assume a situation in which the cognitive relay is close to the SU RX and we have E[\textbar \textit{$\gamma $${}^{(sr)}$}\textbar ${}^{2}$]=3, E[\textbar \textit{$\gamma $${}^{(ss)}$}\textbar ${}^{2}$]=3 and E[\textbar \textit{$\gamma $${}^{(rs)}$}\textbar ${}^{2}$]=8. Both of these figures show optimality of the proposed schemes. However, comparing Figs 3, 8 and 9, we see that the throughput capacity is lower if the cognitive relay is located close to either the SU transmitter or receiver.
\section{Conclusion and Future Works}
In this paper, we optimized the throughput of a CRN with cognitive relaying by joint spectrum sensing and resource allocation under OFDM transmission. An optimization problem was formulated and solved, while meeting constraints such as the interference power introduced by the SU and cognitive relay to the primary system, subcarrier pairing, the miss-detection and false alarm probabilities in each CR subcarrier. By solving the optimization problem, a new iterative algorithm was proposed to obtain energy detector thresholds, subcarrier pairing and the transmit power levels of the SU and the cognitive relay. Moreover, an alternate low complexity, suboptimal scheme was also proposed, which achieves a performance close to the main proposed scheme according to the simulations. Simulation results illustrate that the proposed schemes achieve significantly higher throughput capacity and achievable total rate than those of existing schemes, such as fixed subcarrier pairing, the scheme based on initial spectrum sensing results and the classical scheme without use of the cognitive relay, as presented quantitatively in Section 6. 

The extension of this work can be considered for multiple relays with considering relay selection. Then, the work should be on the cognitive multi-relay network environment while joint optimization of power allocation, subcarrier allocation, subcarrier pairing and relay selection is considered. Furthermore, the purpose is to maximize the weighted sum rate of the network under mutual interference constraints. 

\begin{figure}
	\centering
		\includegraphics[width=0.5\textwidth]{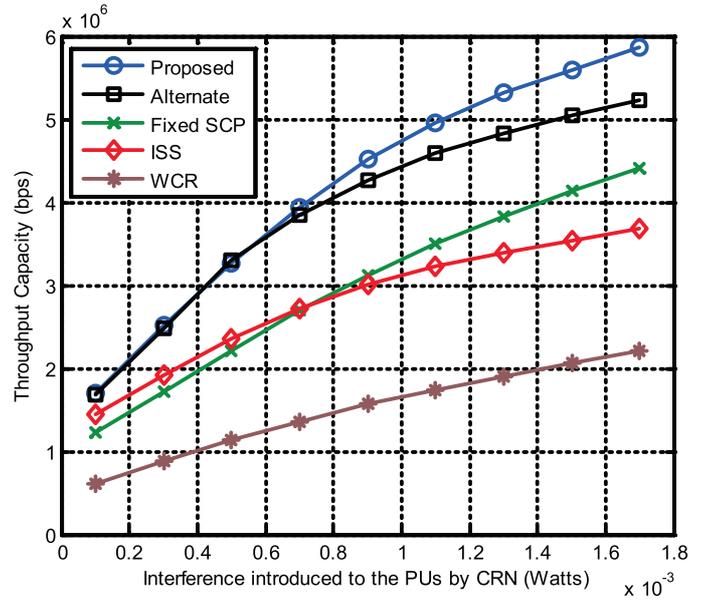}
\caption{\footnotesize{Throughput capacity of the CRN versus interference introduced to the PUs by several schemes with $E[|\gamma^{(sr)} |^2]=3$, $E[|\gamma^{(ss)} |^2]=3$ and $E[|\gamma^{(rs)} |^2]=8$}}
\end{figure}

\section*{Appendix}
\centerline{Verifying the Convexity of Optimization Problem}
For the objective function which is expressed in \eqref{GrindEQ__12_}, If $\nabla ^{2} C<0$, then C is convex and its Hessian is not positive semi-definite. We rewrite the objective function as
\begin{equation} \label{GrindEQ__38_} 
C_{i,j} =f_{1} (P_{i,j} )\times f_{2} (\lambda _{i} ,\lambda _{j} ), 
\end{equation} 
where
\begin{equation} \label{GrindEQ__39_} 
f_{1} (P_{i,j} )=\log _{2} (1+\theta _{i,j} P_{i,j} ), 
\end{equation} 
\begin{equation} \label{GrindEQ__40_} 
f_{2} (\lambda _{i} ,\lambda _{j} )=(1-Q(\frac{\lambda _{i} -M\sigma _{n}^{2} }{\sqrt{2M} \sigma _{n}^{2} } ))\times (1-Q(\frac{\lambda _{j} -M\sigma _{n}^{2} }{\sqrt{2M} \sigma _{n}^{2} } )), 
\end{equation} 
while $\theta _{i,j} =\frac{\gamma _{i,j} }{q_{i,j}}$. As shown in [26], to provide the concavity for $f_{1}$, we have $\gamma _{i,j} P_{i,j} >1.7183$ and by following Eqn \eqref{GrindEQ__17_}, $\nu_{i,j}$=1.7183$\gamma_{i,j}$. Also, Hessian determinant of $f_{2}$ is
\begin{equation} \label{GrindEQ__42_} 
\begin{array}{l}
{\Delta _{H_{2} } =f_{2}^{''} (\lambda _{i} )f_{2} (\lambda _{j} )f_{2} (\lambda _{i} )f_{2}^{''} (\lambda _{j} )-(f_{2}^{'} (\lambda _{i} )f_{2}^{'} (\lambda _{j} ))^{2} } \\ {{\rm \; \; \; \; \; \; }=\frac{1}{2\pi } x_{i} x_{j} e^{-\frac{x_{i}^{2} +x_{j}^{2} }{2} } -(\frac{1}{2\pi } e^{-\frac{x_{i}^{2} +x_{j}^{2} }{2} } )^{2} } \\ {\; \; \; \; \; \;} =\frac{1}{2\pi } x_{i} x_{j} e^{-\frac{x_{i}^{2} +x_{j}^{2} }{2} } \{ [1-Q(x_{i} )][1-Q(x_{j} )]\\ {{\rm \;  \;  \; \; \; \;\; \; } -\frac{1}{2\pi x_{i} x_{j} } e^{-\frac{x_{i}^{2} +x_{j}^{2} }{2} } \} .} \end{array} 
\end{equation} 
We assume that \textit{x${}_{i}$} and \textit{x${}_{j}$} are positive and so the first term in RHS of Eqn \eqref{GrindEQ__42_} is positive. Since in this scenario, we want to obtain the values of energy levels which\textit{ }are derived from the same where, we assign\textit{ x${}_{i}$}=\textit{x${}_{j}$}. This means that we jointly restrict the value of energy detector thresholds in the SU TX and cognitive relay, while in the case where both of them apply one CR subcarrier $\{ [1-Q(x_{i} )]^{2} -\frac{1}{2\pi x_{i} ^{2} } e^{-x_{i}^{2} } \} $ is increasing, and so we can set the possible least value for \textit{x${}_{i}$${}_{ }$} to satisfy positivity of the Hessian, which is obtained\textit{ }by\textit{ x${}_{i}$=}0.507. So, $p_{f} (\lambda _{i} )\le \beta _{i} =0.3061.$

\noindent We see that the result of this is approximately the same as the one in [26], which considers a CR with no relaying. So, by the obtained values for $\beta _{i} {\rm \; and\; }\nu _{i,j} $ the concavity of the objective function is achieved. Also, for Eqn \eqref{GrindEQ__13_}, if \textit{$\alpha_j$} = 0.5, since \textit{Q}(\textit{x})\textit{ }is convex, then this constraint is convex.
We see that both of the obtained values for $\alpha_j$ and $\beta_i$ ($\forall i, j$) are reasonable bounds in practice. 

\section*{Acknowledgment}
This work was supported in part by NSF \#0917008 and \#0916180.

\end{document}